\documentclass[conference]{IEEEtran}
\ifCLASSINFOpdf
\else
\fi

\usepackage{url}
\usepackage{graphicx}
\graphicspath{{figs/}}
\DeclareGraphicsExtensions{.pdf,.jpeg,.png,.jpg}

\begin{document}

\bstctlcite{IEEEexample:BSTcontrol}

%
\title{{ERM}rest: an entity-relationship data storage service for web-based, data-oriented collaboration.}

\author{\IEEEauthorblockN{Karl Czajkowski, Carl Kesselman, Robert Schuler, Hongsuda Tangmunarunkit}
\IEEEauthorblockA{Information Sciences Institute\\Viterbi School of Engineering\\
University of Southern California\\
Marina del Rey, CA 90292\\
Email: \{karlcz,carl,schuler,hongsuda\}@isi.edu}
}



\maketitle

\begin{abstract}

Scientific discovery is increasingly dependent on a scientist's
ability to acquire, curate, integrate, analyze, and share large and
diverse collections of data.  While the details vary from domain to
domain, these data often consist of diverse digital assets (e.g. image
files, sequence data, or simulation outputs) that are organized with
complex relationships and context which may evolve over the course of
an investigation. In addition, discovery is often collaborative, such
that sharing of the data and its organizational context is highly
desirable. Common systems for managing file or asset metadata hide
their inherent relational structures, while traditional relational
database systems do not extend to the distributed collaborative
environment often seen in scientific investigations. To address these
issues, we introduce ERMrest, a collaborative data management service
which allows general entity-relationship modeling of metadata
manipulated by RESTful access methods. We present the design criteria,
architecture, and service implementation, as well as describe an
ecosystem of tools and services that we have created to integrate
metadata into an end-to-end scientific data life cycle. ERMrest has
been deployed to hundreds of users across multiple scientific research
communities and projects. We present two representative use cases: an
international consortium and an early-phase, multidisciplinary
research project.

\end{abstract}


%
\IEEEpeerreviewmaketitle

\section{Introduction}

Scientific discovery has undergone a profound transformation, driven
by exponentially increasing amounts of data generated by
high-throughput instruments, pervasive sensor networks, and
large-scale computational analyses~\cite{bell2009beyond}.  Scientific
collaboration has always been {\em data-oriented}, enabled by the
exchange of data through the lifetime of an investigation. With todays
data deluge, the traditional methods for data organization and
exchange during a scientific investigation are inadequate. This
results in significant investigator
overheads~\cite{kandel2012enterprise} and unreproducable
data~\cite{Begley2013}.  While much attention has been given to the
publication, citation, and access of {\em curated} scientific data
intended for publication~\cite{borgman2012conundrum}, little attention
has been paid to the data management needs which show up in daily
practice of data-rich scientific collaborations.

In current practice, scientists often rely on directory structures in
shared file systems, with data characteristics (i.e. metadata) coded
in file names, text files, or spreadsheets.  If collaboration extends
beyond one institutional boundary, the data may be stored in a cloud
based storage system, such as as Dropbox or Google Drive.  These
approaches are error prone, become fragile as the complexity or size
of the data grows, are hard to evolve over time, and make it difficult
to search for specific data values.

In previous work, we have argued for an alternative approach based on
{\em scientific asset management}~\cite{deriva-escience}.  We separate
the ``science data'' (e.g. microscope images, sequence data, flow
cytometry data) from the ``metadata'' (e.g. references, provenance,
properties, and contextual relationships). We have also defined a
data-oriented architecture which expresses collaboration as the
manipulation of shared data resources housed in complementary object
(asset) and relational (metadata) stores~\cite{DOA}. The metadata
encode not only properties and references of individual assets, but
relationships among assets and other domain-specific elements such as
experiments, protocol events, and materials.

A relational metadata model can be expressed as an {\em
  entity-relationship model}~(ERM), defining entity types (tables) and
relationships (foreign keys and associations). Many projects can be
well-served by simple models with only a handful of entities and
relationships, and non-experts can easily think about their domain in
terms of the main concepts which they want to manipulate [14]. We
introduce {\em Entity-Relationship Models via Representational State
  Transfer}~(ERMrest), a web service where ER models can be created
and maintained by clients---incrementally introducing, using, and
refining domain concepts in a collaborative, mutable metadata
store. The service provides a full-featured
RESTful~\cite{fielding2002principled} interface to underlying
relational data and models.

Despite the popularity and utility of relational databases, and
renewed focus on their Cloud hosting, their data often remain locked
behind application-specific services and treated as an
internal component. We argue that relational storage should be made
directly accessible to web clients for scientific collaboration.
Costly, application-specific services to control data access and
update are replaced with generic storage tier rules for atomicity and
data integrity, combined with fine-grained authorization to enforce
trust-based chains of custody within data-sharing communities. Thus,
end users and user-agents can directly consume and contribute science
data within a flexible and adaptive collaboration environment.

In this paper, we focus on the requirements, design and implementation
of the ERMrest service. We draw upon our experience with several
data-sharing projects to define the problem space, but report on two
representative applications: first, a large complex, multinational
consortium accelerating research on G-Coupled Protein Receptors; and
second, a multidisciplinary synaptomics research project as an example
of an early-phase, exploratory collaboration.

The rest of this paper is organized as follows.  In
Section~\ref{section:characteristics} we discuss key characteristics
of the data-oriented collaboration problem domain.  In
Section~\ref{section:ermrest}, we describe in detail the ERMrest
service, followed in Section~\ref{section:ecosystem} by a brief
discussion of the other components in the ERMrest software
ecosystem. In Section~\ref{section:gpcr}, we describe the GPCR and
synaptomics use cases in more detail. We conclude with related work in
Section~\ref{section:related} and conclusions in Section~\ref{section:conclusions}.

\section{Application Characteristics and Challenges}
\label{section:characteristics}

Our perspective on scientific asset management, data-oriented
collaboration, and the specifics of ERMrest as a metadata catalog
service have been informed by a number of scientific projects for
which we provide bioinformatics support. Rather than bespoke solutions
for each project, we have sought archetypal requirements and evidence
of feature gaps where our model-driven, data-oriented tools could be
enhanced to support broad classes of collaboration.

Here, we outline the main application characteristics, illustrated
by five ongoing projects:
A)~the hub for the FaceBase project, organizing a central repository
for data generated by a number of individually-funded spoke sites;
B)~the microscopy core for the Center for Regenerative Medicine and
Stem Cell Research~(CIRM), offering microscope slide-scanning as a
service;
C)~the GUDMAP project, curating microscope imagery with assessments
and annotations by domain-experts;
D)~the GPCR~Consortium, an international collaboration to discover
and analyze G-Protein Coupled Receptor molecular structures;
and finally,
E)~Mapping the Dynamic Synaptome, a multidisciplinary effort to
develop methods for in vivo measurement of the synaptome.

\subsubsection{Heterogeneous metadata}

A model-neutral metadata catalog and associated services and
applications allow reuse of the same technology without repeated
software development to adapt services and applications for each
project.  Across our projects, there are multiple data types and
formats generated on a daily basis, e.g. DNA sequencing; flow
cytometry; chromatography; multi-dimensional simulation results; 2D
and 3D microscopy including tiled, multi-resolution zoom pyramids; 3D
CT and micro-CT; and 2D time-series (video). These assets must be
acquired, named, stored, and tracked with scientific context and
provenance metadata so that they can be found and consumed by relevant
down-stream science users and processing pipelines.

As file-like objects, all assets can share generic object metadata
concepts such as reference, size, checksums, or file type. However,
different formats and modalities may have wildly differing metadata
relevant to consumers, e.g. different kinds of timestamps,
different kinds of instrument identifier and acquisition parameters,
and different dimensional or shaping information.

Even more significant than variation in asset properties, different
domain models can vary in their encoded relationships and non-asset
entity types. These additional ERM elements can provide significant
scientific context to the underlying assets, e.g. recording
information about events, protocols, and materials. Depending on the
level of formality in projects, different kinds of provenance and
quality-control metadata may also be introduced.

\subsubsection{Evolving data models}

Not only should the technology be configurable for different
project-specific models, but the models should be allowed to continue
to evolve throughout a project's life-cycle, i.e. while the system is
in use.  Early-phase, exploratory projects need quick setup and simple
models while users establish experiment protocols, collaboration
methods, metadata nomenclature, and coding standards. As projects
mature, users may identify new use cases and new formalisms which
early users would have never prioritized. Finally, as projects
transition to curating published data products, the modeling goals can
drift further from the lab or process-management goals of an active,
experimental project.

As an example, the GPCR scientific workflow is complex, involving many
distinct experiment methods with different data-handling
pipelines. Our pilot model focused on core domain concepts and a
proof-of-concept pipeline, introducing the system for early test
users. This core model continues to evolve with experience, similar to
most of our other projects; but, due to the breadth of activities
within the consortium, it also expands in bursts as more tasks and
objectives are incorporated into the data-management system.

Typical of many repositories and analysis-based projects, the FaceBase
hub metadata was initially modeled after a bulk export from an
ancestor project. This repository model expands as new spoke data is
integrated. Likewise, the CIRM and GUDMAP models were initially
informed by existing image archives and microscopy idioms. In all
projects, models continue to be refined in response to evolving user
needs, newly identified search goals, and various metadata curation
objectives.

The Synaptome project started from scratch with little guiding
data. We quickly found that very simple models could be directly
aligned to nascent laboratory workflow steps and tuned as we
worked. The catalog in some ways embodies a structured laboratory
notebook, providing data-entry rules where we might otherwise face
creative chaos.

\subsubsection{Heterogeneous data source integration}

Data and metadata may be sourced from legacy data sets, exports, and
publications; specialized instruments and instrument control software;
third-party laboratories; existing databases and lab information
management systems; or loosely coupled sources like lab notebooks,
spreadsheets, text files, and manual user entry. To effectively
support collaboration, data-management solutions must lower barriers
to data entry and encourage the collection of experiment metadata
before important scientific context is lost and the value of the data
extinguished.

Budding projects often have only spreadsheets or other text manifest
files, often with enigmatic naming schemes and formatting. As they
mature, these projects demand more structured methods for recurring
data acquisition.  To integrate a variety of instruments and other
repeatable processes, we observe the need for both custom scripting
and easily reusable file-handling tools discussed later in
Section~\ref{section:ecosystem}. To facilitate manual procedures that
would otherwise be covered by personal laboratory notebooks, we
observe the need for simple data-entry tools to prompt the user for
structured information and provide immediate data-validation feedback.
These tools should be driven by the project's evolving data models so
that new model elements are automatically fitted with basic data-entry
forms and procedures.

GPCR has three academic sites all having existing local databases
containing construct design and expression information which must be
extracted, transformed and merged in the consortium catalog. FaceBase
inherited a relational database dump from a content-management system
used to construct the previous, bespoke hub website, and continues to
receive data submissions as smaller exports from spoke sites. CIRM and
GUDMAP began with many image files but very little structured
metadata; they instead acquire most metadata as data-entry by
interactive users.

\subsubsection{Data processing pipeline integration}

Aside from data sources, which introduce data and metadata into the
system based on human activity and other external laboratory events,
there is also a recurring need to integrate data processing
pipelines. These consume existing data and produce derived data and/or
metadata which return to the shared data store to enrich the
collaboration. There is a wide spectrum of pipelines, varying in terms
of: technology dependencies; cost and duration of
processing; level of human input in triggering or controlling the
pipeline; number of intermediate results which are captured back
into the shared asset and metadata stores; and semantic
relationship of any derived results to the previously existing shared
data. We observe the need for an environment which is not biased
towards any one form of processing pipeline, so that a variety of lab
processes and technologies can easily contribute to a managed data
collaboration.

\subsubsection{Data discovery, access, and consumption}

Ultimately, shared science data should be searchable and accessible
for data exploration. A user interface is needed to provide rich
searching and browsing capability to any user with a web browser. It
should guide users through the diverse data types available during all
stages of complex scientific workflows and during all phases of
projects. Like data-entry tools, this capability should be
model-driven and provide general insight into the current state of the
shared data resources even as project models change, rather than being
narrowly focused on one specific scientific workflow or model.

Depending on the data type and nature of collaboration, different
modes of data consumption are important: general metadata summary
tables and documents; coded quality assessments; statistical
summaries; plots, thumbnails, and online previews; or download links
and URLs usable with domain-specific applications and dedicated
workstations. We observe a need to allow optional, project-specific
tailoring of these presentations beyond what a pure ER domain model
can express.

\subsubsection{Differentiated access control}

In order to address a range of projects, configurable policies are
needed to allow different mixtures of public, authenticated but
read-only, and authenticated read-write access. These different
policies must be applicable in a fine-grained way, so that multiple
classes of assets and metadata can be managed in one project and given
different access policies.

As a stable repository resource, the FaceBase hub receives data
submissions from contributing spokes. The hub curates and integrates
these disparate data products to provide three tiers of repository
data access: public metadata and thumbnail imagery to advertise the
data resource; login-protected online data requiring acceptance of a
data-use agreement; and offline human subjects data protected by IRB
and strict data distribution methods.

The CIRM microscopy core receives physical specimens from client
biologists and provides a slide scanning service, making resulting
digital imagery available to each corresponding slide owner. The core
staff manage this ongoing acquisition process, curating metadata and
bulk storage. The ongoing GUDMAP collaboration studies a predefined
set of images and produces micro-anatomy annotations and other curated
byproducts, giving different levels of read or read-write access to
images, image metadata, and other byproducts depending on user role.

The GPCR data are generated from three different sites in disparate
locations. Only the lab members associated with the site performing
the experiment are allowed to create or update subsequent data along
the experimental workflow. In addition to data sharing among the three
sites, a subset of data must be shared across the distributed
consortium as well as the broader scientific community, all according
to the consortium data sharing policy and user roles.

The Synapse project involves a small, multidisciplinary team including
individuals from several labs. These few individual members have full
access to embargoed data.

\section{ERMrest}
\label{section:ermrest}

To address these application requirements, ERMrest provides a
relational metadata store as a web service. It exposes multiple {\em
  catalogs}, each with its own {\em access-control} lists (ACLs), {\em
  ER model}, and {\em content} following that model. Unlike many
data-management solutions which are designed or deployed with {\em a
  priori} knowledge of the data model, ERMrest continuously adapts to
the catalog model, providing model-driven interfaces.\footnote{The
  current ERMrest research software, based on PostgreSQL, understands
  many common SQL data definition concepts. But, it is not capable of
  handling completely arbitrary schemas using the full PostgreSQL
  object-relational repertoire.}  With this approach, we eliminate
redundancies in data modeling and data model seepage at multiple
levels of the traditional Web application stack. We enable users to
create and evolve data models which represent the semantic concepts in
their domain, without the typical slow cycle of product updates
involving user feedback, new development, and database schema
migrations.

Our interface design approach can be summed up as ``keep simple things
simple, and make complex things possible.'' We want to support idioms
common to web services, clients, scripts, and programming models. We
prefer a rich set of web resources, over which conventional HTTP
methods can be used from even trivial clients, using straightforward
content representations such as Javascript Object Notation (JSON) and
comma-separated values (CSV).

\subsection{Technical Goals and Scoping}

\subsubsection{Meaningful URLs}

We consider it acceptable or even desirable for client programmers to
induce new URLs based on an understanding of the server's resource
space. We do not expect our typical clients to exhaustively crawl a
connected set of representations to find every URL they might need to
visit.  However, we also recognize the value of linked data and
consider it useful and important for URLs to be created, discovered,
exchanged, stored, and later accessed. We aim to have well-behaving
URLs to reference ERMrest resources, and domain-specific ERMs may of
course embed URLs in their entity data. When ERM concepts or content
are encoded in URLs, we want simple, human-readable URL formats to
facilitate developer and user comprehension. However, we also want
rigorous formats to handle arbitrary concept and data values
which may include challenges such as punctuation, whitespace, and
non-ASCII characters.

\subsubsection{Collaborative Data Architecture}

We aim to directly connect users and user-agents to mutable storage
services without the interposition of bespoke application servers. To
this end, the individual storage services must understand and enforce
fine-grained authorization directly in terms of end user identity,
attributes such as group membership, and access control policies
associated with the managed data resources. Simultaneously, clients
must be aware of the basic rules of engagement for collaborative
storage and tolerate storage state configurations which can be
produced by other members of the collaboration.

\subsubsection{Long-tail Scalability}

Our target audience is not mass consumer applications where hundreds
of thousands to millions of clients need access to the same data sets
or database.  Rather, our audience straddles the ``long tail'' of
e-Science~\cite{borgman2016}, where many small collaborations may each
involve merely dozens of data-producing clients, dozens to hundreds of
data-consuming clients during the active phases of research, and an
unknown number of casual or single-use data-consuming clients in
later, passive phases of scientific libraries and archives. It can be
said that much science data is ``written once and read never,'' but
the value is in being able to find the subsets of data worth
revisiting, and this will rarely be known at the time of
acquisition.

Therefore, where ERMrest needs scaling is in the number of small,
project-specific catalogs, each with a viable content model that
simplifies collaboration and data discovery for that team. It is not
geared to support massive catalogs where the bulk of science data is
the catalog content itself, i.e. large-scale measurement data encoded
directly in relational tables, or where the relational data requires
non-trivial statistical analysis.  To serve the long tail of projects,
we require the same software stack to support many projects without a
developer modifying the service as each project's content model
evolves.

\begin{figure}
  \includegraphics[trim={0 3in 0 2.75in},clip,width=\linewidth]{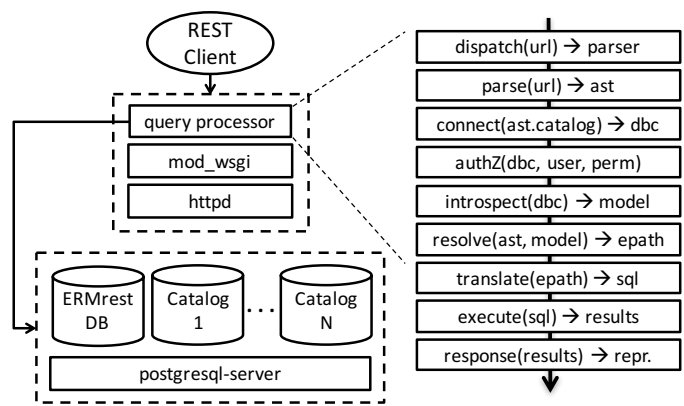}
  \caption{ERMrest service is a model-driven web service implementing
    multi-tenant catalogs as separate PostgreSQL databases, each with
    a domain-specific entity-relationship model. RESTful interfaces
    support catalog, model, ACL, and data creation, retrieval, update,
    and deletion.}
  \label{fig:ermrest_arch}
\end{figure}

\subsubsection{Full Lifecycle Support}

We intend to support the full lifecycle of scientific data including
early experiment design; early and production data acquisition; ad hoc
and repeated analyses; and publication. This flexibility demands:
configurable access controls; an ERM which can evolve throughout the
project lifetime; and rich content access interfaces, capable of
supporting incremental update and retrieval as well as bulk search. A
single catalog should be able to support a mixed load of ongoing
research, including embargoed data, while also exposing final data
to a wider audience. However, it should also be
practical to split data into separate catalogs or service instances or
to export data to publishing systems.

\subsubsection{Data Portability}

We also recognize that projects often have changes in direction,
funding, and priorities which force technology change. We therefore
wish to mitigate the scientists' risk and worry that data become
captive to a system in need of replacement. Through the use of
standard ERM concepts and standard tabular data encodings, we can
ensure that raw data is easily exported via the web interface. While
implementing client-visible ERM concepts in our service, we also wish
to allow a project administrator to intervene directly in the backing
data store.  This can allow one to export the full catalog (including
model); to bypass limitations of our service interface for more exotic
data access modes; or to customize the catalog ERM in ways not
possible through the current web resources. We consider such data
portability to be essential for long duration stewardship, and
therefore a critical part of any scientific data management
methodology.

\subsection{HTTP Interface and Semantics}

We use an attribute-based naming style to structure URLs in our
interface. This means that domain concepts such as table names or
entity keys will appear in URLs. We also follow a strict reading of the
URL encoding rules~\cite{RFC3986}.  We use {\em reserved characters}
(mostly punctuation characters) as syntax in the URL. We require these
special characters to be ``percent-encoded'' when appearing as regular
data not meant as ERMrest syntax. This permits arbitrary Unicode
content within atomic elements of the URL, yet reads very simply for
common instances using plain ASCII content. We also support
complementary web standards including content-negotiation and
opportunistic concurrency control.

\subsubsection{Catalog Management}

To support multi-tenancy with differentiated access control, we
expose catalog management as a top-level resource {\tt
  /ermrest/catalog/} where a POST method can create a new catalog.
Each new catalog has a brief document representation and its own
URL, e.g. {\tt /ermrest/catalog/1}, supporting the DELETE operation to
retire it. The catalog document includes its current ACLs, which are
also exposed as sub-resources, each having their own URL for direct
management.

\subsubsection{ERM Management}

To support domain models, the catalog ERM is exposed as another
hierarchical document structure at a sub-resource inside each catalog,
e.g. {\tt /ermrest/catalog/1/schema}. A number of kinds of
sub-resource (each instance having its own URL) are used to
manipulate the ERM incrementally:
\begin{description}
\item[Schema] A namespace within a catalog.
\item[Table] A table defined within one schema.
\item[Column] A column defined within one table.
\item[Key] A uniqueness constraint on one table.
\item[Foreign Key Reference]~ \hfill

  An ``outbound'' reference constraint from one table to a key in the
  same catalog.
\item[Comment]~ \hfill

  A short, human-readable string can document various model elements.
\item[Annotation]~ \hfill

  Machine-readable documents can augment various model elements for
  semantic enhancement.
\end{description}
All model resources support GET, while very few support PUT for
mutation. Instead, resources support DELETE to prune out elements and
containers support POST to introduce new sub-resources.  Entire
sub-trees of resources can be created en masse, e.g.  a table with all
its columns and constraints or a schema with multiple tables can be
sent in a single POST.

New tables are always specified without data content, and separate
resources must then be manipulated to load entity data. Deletion of a
table or column from the ERM also causes all of its associated data
content to disappear.

\subsubsection{ERM Annotations}
\label{section:annotations}
Our model annotation mechanism allows semantic enhancement of pure ERM
concepts. Schemas, tables, columns, keys, and foreign key references
each bear an annotation container. The payload documents are keyed to
distinguish different kinds of annotation on the same model element,
using URIs to manage key collisions.  An annotation is a statement
about the annotated model element, and the key and payload are akin to
a predicate and object, respectively. Anyone may invent new kinds of
annotation and assign a key URI using any URI naming scheme for which
they have naming authority.  In practice, we specify small,
single-purpose annotations using {\tt tag:} URIs~\cite{RFC4151} and
each defines very simple payload which our custom clients can pick and
choose to support.

The payload must be a valid JSON document which could be as simple as
{\tt null}.  A client may interpret any annotation they recognize and
should ignore any they do not understand or which they do not care to
observe.  ERMrest does not interpret annotation payload but merely
stores and distributes them to help coordinate clients wishing a
shared understanding beyond the purely structural ERM rules.

\subsubsection{Catalog Content}
To support rich data access for both incremental access and bulk
search, the catalog data content is exposed through a set of {\em
  access mappings} which apply to the same catalog under different URL
prefixes. Each access mapping defines a particular attribute-based
naming syntax for URLs, naming a family of data resources. The
different access mappings provide overlapping access to the same
mutable catalog store, in essence providing aliased URLs suitable for
different client use cases.

Each ERMrest data URL denotes a {\em tabular data set}. Each URL
selects the catalog; selects the access mapping; {\em navigates}
tables and foreign keys of the ERM as a join product; sets a {\em
  target table} and {\em projection}; optionally indicates {\em data
  filters}; and optionally indicates {\em page position}. These
concepts are explored in more detail below.

\subsubsection{Navigation and URL Structure}

A path-like notation is used in data URLs to express joins and filters
as a kind of ``drill-down'' from one table to another. Consider a few
illustrations (each extends the previous when appended):
\begin{description}
\item[{\tt /ermrest/catalog/1/entity/Subject}]~ \hfill

  The entities from the {\tt Subject} table;
\item[...{\tt /id=17}]~ \hfill

  ...but only where {\tt id} is {\tt 17};
\item[...{\tt /Image}]~ \hfill

  ...instead returning entities from {\tt Image} which are related to
  the {\tt Subject} entity by foreign key;
\item[...{\tt /acquired::gt::2016-01-28}]~ \hfill

  ...but only where {\tt acquired} date is more recent than
  2016-01-28;
\item[...{\tt @sort(acquired::desc::,id)}]~ \hfill

  ...sorted by {\tt acquired} date (descending), with {\tt id} to
  break ties;
\item[...{\tt @after(2016-02-24,15)}]~ \hfill

  ...only for records following stream position ({\tt 2016-02-24},
  {\tt 15}) for paging;
\item[...{\tt ?limit=20}]~ \hfill

  ...limit the page to 20 entities;
\item[...{\tt \&accept=csv}]~ \hfill

  ...demand CSV output, e.g. for bookmarks.
\end{description}
The preceding entity-set with join, filters, and page position would
then have this complete URL (wrapped to fit):
\begin{verbatim}
/ermrest/catalog/1/entity/Subject/id=17\
/Image/acquired::gt::2016-01-28@sort(ac\
quired::desc::,id)@after(2016-02-24,15)\
?limit=20&accept=csv
\end{verbatim}
Because the {\tt entity} API provides a whole-entity resource mapping,
no projection syntax is included, and the denoted set will
include all columns of the {\tt Image} target table.

\subsubsection{Filter Predicates}

To support both bulk search and simple incremental row access, 
data URLs can express filters where each filter predicate can compare
a column instance to a constant value.  We support equality and
inequality operators, an is-null operator, and several text-pattern
operators.\footnote{We expose underlying database primitives for {\em
    regular-expression} matching, which also supports substring match,
  and {\em text search vectors} which use a natural language parser
  and dictionary for word-stemming and synonym resolution. We
also expose a special wildcard column name {\tt *} which allows
text-pattern operators to search all text-like columns of an entity at
once, rather than having to express disjunctive predicates repeating
the same operator and constant pattern for many columns.}

Negation, nesting parentheses, and both disjunctive and conjunctive
logical combiners allow arbitrary boolean functions to be expressed
as denormalized expression trees or conjunctive/disjunctive
normal form.  The filter function is evaluated over the path as a
whole, so as with SQL, each member of the join product must satisfy
the entire filter expression to avoid being excluded from results.

\subsubsection{Structural Search}

To support bulk searches, the path-like syntax introduced above can be
used to join tables in a chain of pair-wise joins. At each stage, the
preceding {\em path context} is the table instance used for resolving
filters or further joins. The final path context is the target table
for projection in retrieval methods or determining the target of
mutation methods. Using more elaborate URL syntax, more complex
join-tree expressions permit {\em structural search}, where the
denoted entities are filtered by a sprawling join pattern which
reaches into different parts of the ERM simultaneously. This more
explicit join syntax allows one to choose among multiple alternative
foreign key reference constraints by calling out their key or foreign
key endpoints; to configure outer join variants instead of the default
inner join; and to join new table instances to any ancestor of the
path rather than only the immediate parent element.

As a result, many ``conjunctive'' or ``select-project-join'' queries
can be encoded directly as data URLs. However, we do not allow
arbitrary join conditions, limiting the user to inner and outer joins
based on navigation of foreign key reference constraints in the ERM.

\subsubsection{Projection and Aggregation}

The {\tt entity} mapping has implicit projection of all columns in
target table. Other mappings require explicit projection clauses at
the end of the path. These clauses express a list of column instances,
with optional syntax for: assigning an external alias to rename
columns in the external representation; qualifying column names with a
table instance alias to project columns from any element of the query
path other than the final path context; or aggregate functions to
compute simple value reductions like counts, min/max, or array
aggregation.  As a convenience, bare columns may also be used in an
aggregate projection context, in which case a representative value is
arbitrarily chosen from one of the rows prior to reduction. For
grouped aggregations, we split the projection into two parts: the
group-key columns and the optional columns subject to aggregate
reduction or aggregate update.

\subsubsection{Mutation not Quite RESTful}

We aim to avoid the ``anti-pattern'' of remote-procedure
call over HTTP, where arbitrarily stateful operations are tunneled
through an opaque service endpoint.  However, providing pragmatic,
fine-grained access to relational storage requires some compromise. It
is impractical to retrieve, edit, and submit entire tables full of
data for every update; it is likewise impractical to require one HTTP
request per row for actions involving many rows.

Every tabular data set may be retrieved or deleted using the GET and
DELETE methods, respectively. Only a small subset of possible data
URLs support PUT or POST methods to apply new content. Our mutation
methods are not purely RESTful. They each manipulate a set of rows in
the underlying ERM storage through the access mapping, and
side-effects are visible through every data URL which overlaps the
targeted content. It is perhaps more accurate to think of each data
URL as representing a ``query'' resource, the results of which can be
retrieved or used as input to a server-side mutation operation which
may merge further data supplied as input from the client.

Data URLs do not transition to ``not found'' status after
deletion, but merely begin to denote the empty set after their
matching entities or attributes are deleted.  Our PUT, POST, and
DELETE methods correspond more accurately to a PATCH method with
different flavors of patch instruction, merging some user-designated
set into the relational store.  All operations express bulk,
tabular access rather than having one HTTP request per target
entity. We think these patch-based access methods are important for
any web-based relational store, but we remain open to reformulating
their surface syntax to better conform to web standards in the future.

\subsection{Relational Data Mappings}

For simplicity, ERMrest always uses data formats capable of
representing a sequence of tuples, even if the particular named data
resource is a degenerate case with set cardinality of one (single
tuple) or zero (empty set). Because we always return sequences, an
empty set is considered valid and will be retrieved with a successful
{\tt 200 OK} status code. Simple content-negotiation allows clients to
choose between two main tabular data representations. With {\tt
  application/json}, a JSON array of objects represents a sequence
of tuples with named column values. With {\tt text/csv}, a CSV document
specifies header row and zero or more data rows.  As suggested
previously, we support multiple data access mappings in the URL space
of the catalog:

\subsubsection{{\tt entity}}

Set of whole entities of one table. Supports insertion, retrieval,
update, and deletion of whole entities from that table. For retrieval
and deletion, supports filtering which may include contextual
information from relational joins.

\begin{figure}
  \includegraphics[width=\linewidth]{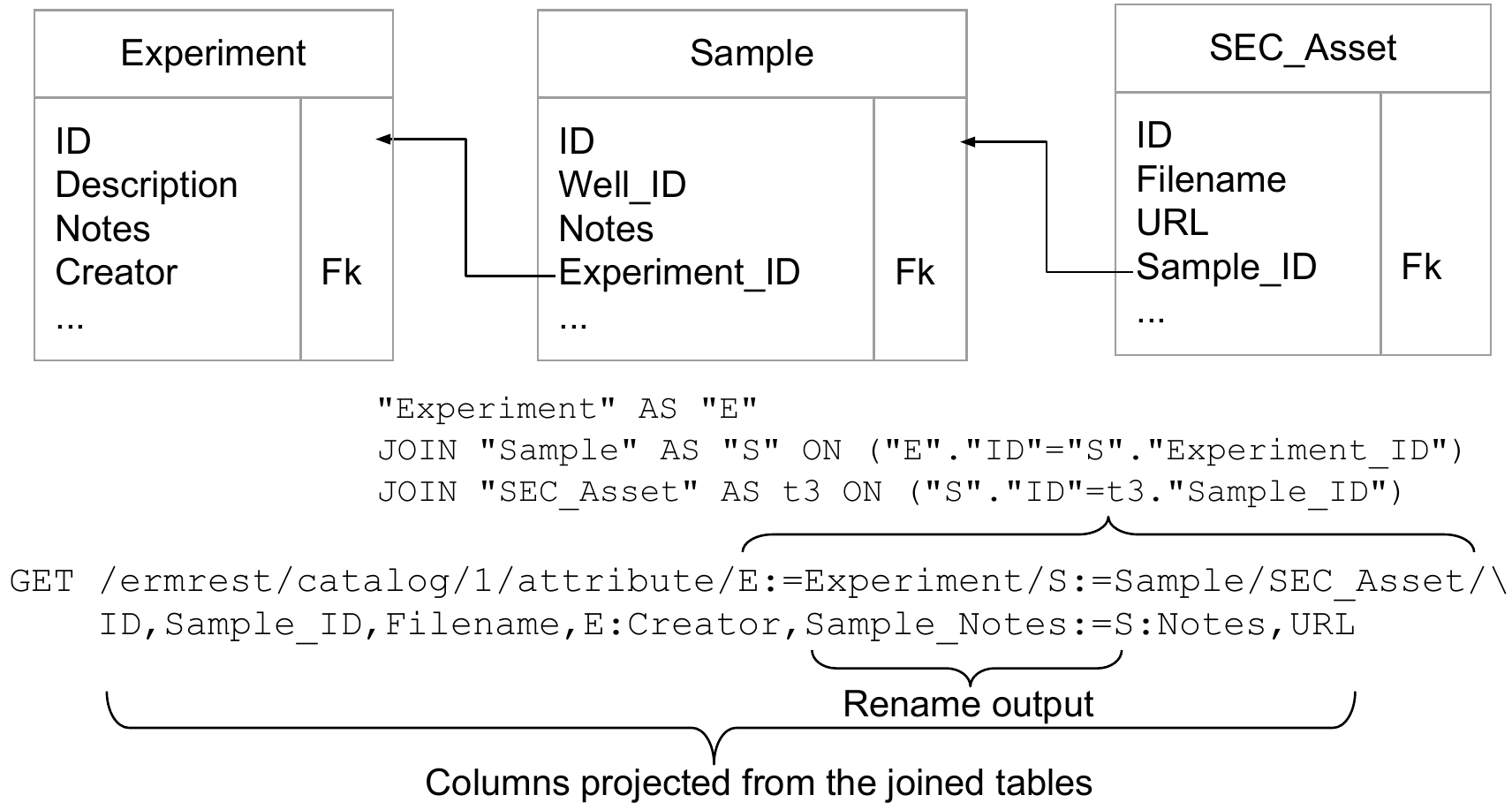}
  \caption{Structured query example with two joins and an {\tt
      attribute} projection including columns from all three tables in
    the depicted entity-relationship model.}
  \label{fig:query_example}
\end{figure}

\subsubsection{{\tt attribute}}

Set of partial entities of one table. Supports retrieval and deletion
of attribute content. Deletion of attributes means the {\em clearing}
of individual columns in existing entities but never inserts nor
removes entities from the target table. Supports filtering which may
include contextual information from relational joins. For retrieval,
the projected content may also include contextual information from the
same relational joins.

\subsubsection{{\tt attributegroup}}

Set of tuples reduced by group aggregation and projected from one
table or join. Supports retrieval and update of grouped attribute
content. For retrieval, supports filtering and projection which may
include contextual information from relational joins. Projections
include group keys and may include computed aggregates. For update,
group keys are used to correlate existing entities in the target table
with tuples in the supplied input representation, and projections are
used to assign further input columns to target columns. Each input
tuple represents one update group where the matching entities in the
target table receive the same column updates.

\subsubsection{{\tt aggregate}}

A set of exactly one tuple reduced and projected from a table or
relational join with or without filtering predicates. The single
member of the set represents the whole aggregate.

\subsection{Data Retrieval Examples}

Using the query-language features described above, we can formulate a
number of example URLs and retrieval scenarios. Consider the
three-element ERM depicted in Figure~\ref{fig:query_example}.  Using
this model, we can retrieve one {\tt Sample} entity:
\begin{verbatim}
GET /ermrest/catalog/1/entity/Sample/ID=5
\end{verbatim}

A more complex retrieval URL is depicted in
Figure~\ref{fig:query_example} itself. Rather than considering only
one table and a constraint on the primary key {\tt ID} column, all
three tables are joined and no filters are present. The query
represents one row per {\tt SEC\_Asset} but projects columns from all
three tables.

Using outer joins, we could find experiments lacking samples:
\begin{verbatim}
GET /ermrest/catalog/1/entity/S:=Sample\
/right(Experiment_ID)/S:ID::null::
\end{verbatim}
which corresponds to this equivalent SQL:
\begin{verbatim}
SELECT t2.*
FROM "Sample" AS "S"
RIGHT OUTER JOIN "Experiment" AS t2
  ON ("S"."Experiment_ID"=t2."ID")
WHERE "S"."ID" IS NULL;
\end{verbatim}

\subsection{Concurrency, Transaction, and Update Model}

Each web request is performed under transaction control and represents
an atomic interaction with the catalog. A client may order a sequence
of requests to different resources and be sure that they have been
committed in that order, but must cope with failures on each
individual request. There is no direct support for multi-request
transactions, two-phase commit protocols, etc.

We synthesize a monotonic catalog version for the HTTP {\tt ETag}
response header and internal model cache management.  The entity tag
is also recognized in {\tt If-Match} and {\tt If-None-Match} headers
to support opportunistic concurrency control.  We use internal
state-tracking tables to know what version the catalog is at. Any
direct, local access to the backing SQL database must inform ERMrest
of changes by invoking special stored procedures which manipulate
these tables.

\subsection{Implementation}

The service is written in the Python language using the web.py
(www.webpy.org) web framework, and is hosted in the Apache web server
via {\tt mod\_wsgi} daemon processes. It uses PostgreSQL databases to
back each catalog. A simple registry tracks all provisioned catalogs
and their backing database connection parameters. We also add a hidden
database schema, {\tt \_ermrest} in each catalog database to store
management state. The system architecture is depicted in
Figure~\ref{fig:ermrest_arch}.

\subsubsection{Access Control}

We integrate with a web-based identity and attribute provider to get
web client identity and group membership.  We implement access control
at the catalog level in the service logic, using ACLs stored in the
management schema within the catalog so that policy, model, and data
are consistently under transaction control. We pass down client
authentication context as trusted SQL session parameters, allowing the
backing database to enforce row-level security policies which can
differentiate individual web users. This policy feature is being
experimentally validated in several pilot projects and will likely be
extended web-based management features in future revisions.

The request processing flow is dispatched to one of two daemon
containers. One handles all ERM-related requests with a SQL access
role which manipulates backing databases and schemas on the fly. The
other handles all bulk data requests with a less privileged SQL access
role. This second role owns SQL views and is granted permission to
access tables but is not their owner. This split-role configuration
enables the desired effect where row-level policies can automatically
filter all stored data access according to web client role.

\subsubsection{Catalog Scalability}

We do not see PostgreSQL as a significant bottleneck. Small catalogs
perform admirably even on very modest server hardware. In our
experience, large data volumes for scientific research usually involve
many, large bulk data assets and require significant investment in
object storage infrastructure. Projects encounter limits to how much
they are willing to spend on terabytes to petabytes of bulk object
storage, while their metadata requirements can still easily be solved
by a commodity server running ERMrest, with room to grow through
basic server upgrades.

At the extreme long-tail end of the spectrum, ERMrest multi-tenancy
allows many catalogs in a single server instance. This allows the
packing of the maximum number of catalogs into the least hardware
resources, while still allowing for an independent data model for each
catalog. At its limits, ERMrest will begin to exhaust the system
resources needed to sustain a pool of active database connections
capable of simultaneously using many database catalogs with low
latency.

However, with its completely open source software stack, the entire
solution is easily virtualized, with one catalog per server. This
approach allows essentially arbitrary scale-out to a world full of
small collaborative projects, where each project can fund its own
quite small instance of the service stack operated either on their own
colocated physical hardware, their own virtualized server platform, or
on cloud hosting.

We operate the majority of our scientific data management projects in
single-project virtual servers, including ERMrest, a static web server
to host GUI applications and static HTML pages, and a companion object
store all behind a single vanity domain name specific to that
project. This approach also benefits our full lifecycle objectives, as
it becomes much easier to prototype, operate, hand-off, or even retire
a project-specific storage environment on a rolling schedule.

\subsubsection{SQL Catalog Customization}

For many projects, we apply some customization to the catalog database
using local SQL mechanisms. ERMrest maps each visible ERM element to a
corresponding SQL element such that local SQL access is complementary
to web-based access. This includes directly applying user-specified
table and column names, column types, and data-integrity
constraints. We exploit this as a vehicle to explore new design ideas
and to evaluate collaboration features before they find their way into
the web service design. This is a {\em trusted interface} not managed
by the same web-based access control policies.

The most common enhancements fall into three main categories: the
application of PostgreSQL-native {\em row-level security} policies;
SQL {\em triggers} to compute certain column values and protect some
as write-once, e.g. for managing accession identifiers; and SQL {\em
  views} to express certain denormalized presentations and group
roll-ups which are generally useful as dashboards for graphical
clients or status representations for workflow clients.

\subsubsection{Exposing SQL Views}

To permit ERM customization, we expose each view as a read-only table
and provide an ERMrest-specific mechanism of {\em pseudo-constraints}
to connect such views into the model, declaring uniqueness or
referential integrity properties which we know the views
obey.\footnote{It is the modeler's responsibility to declare accurate
  pseudo-constraints or encounter confusing query behaviors. The
  system makes no attempt to analyze view definitions and prove the
  correctness of the declared constraints nor to test whether view
  results satisfy the constraints at run-time.}  Thus, our normal ERM
navigation features can extend to such SQL views and all model-driven
client tools can exploit the view. This is an escape-hatch for
problems too awkward to solve otherwise. At the time of writing, such
SQL views require local administrative access to create.

Web-managed views are an interesting topic, both to offer this
ability to less trusted clients and as underpinnings for asynchronous
query interfaces. An asynchronous query would create a pseudonymous
view which can then be accessed page by page using our normal
synchronous retrieval interfaces. Such queries should support
richer query languages and avoid length limitations inherent in
mapping query structure to URL structure. We have deferred
implementing remote view management while we debate expressiveness
versus safety for less skilled or less trusted clients. We continue
to gather use-case requirements to help guide this decision process.

\section{ERMrest Ecosystem}
\label{section:ecosystem}

ERMrest enables an ecosystem of tools and methods.  Here we describe
asset storage, graphical interface, and automated agents which all
contribute to the solution space.

\subsection{Hatrac: Object Store}

Hatrac is simple object-store with the same end-user security
integration used in ERMrest. Hatrac objects are file-like data
elements which are atomically created, destroyed, or updated with new
object-versions. It can use plain filesystem storage or proxy to 
Amazon S3 or compatible object stores. In either case, Hatrac provides
a consistent service interface with end-user authorization for
individual objects or namespace hierarchies.  Hatrac is not directly
dependent on ERMrest and vice versa. However, in all our deployments,
both are typically deployed together to support sharing of assets
and metadata.

\subsection{Chaise: Model-Driven GUI}

Chaise is a suite of graphical user interface applications for browsing,
search, editing, creating, and exporting ERMrest catalog content
via a web browser. Implemented in javascript and HTML5 technology,
Chaise dynamically generates a user interface based on the ERM and
data encountered in the catalog. Chaise provides an experience that
revolves around entities and their relationships. We have explored
{\em faceted search} and basic {\em text search} interfaces, using
structured queries with joins and filter predicates. Search results
are displayed as tabular data, providing navigation to entity-oriented
views which present details of the entity as well as summaries of {\em
  related entities}. A user can browse and explore a linked web of
human-friendly graphical application views isomorphic to the
underlying relational catalog content.

Chaise recognizes a number of optional model annotations to allow the
data modeler to customize presentation.  For example, {\tt
  tag:isrd.isi.edu,2016:visible-columns} supports ordered lists of
columns to show in particular GUI display contexts, overriding the
default model-driven presentation. Other annotations are used to
describe content transforms, where entity data can be interpolated
into an intermediate Markdown~\cite{CommonMark} fragment which is
rendered into HTML and incorporated into the Chaise GUI. For example,
one might render a navigable link using one column's value as anchor
text and another as the destination URL.

\subsection{IOBox: Automated Agents}

IObox (a combined ``inbox'' and ``outbox'') is a family of automated
user-agents for dynamic orchestration of data flow tasks.  A user or
instrument may place new files into an ``outbox'' directory so that
the agent will automatically convey data to ERMrest and/or
Hatrac. Likewise, the agent can retrieve data and leave copies in an
``inbox'' location where a user, instrument, or analysis pipeline can
consume it.

IObox executes an event loop with configured rules. Data come from a
configured event listener or by polling a filesystem location or
catalog query. When data match a configured rule, a chain of handlers
is invoked to process the data.  Built-in handlers can extract
metadata: from file names or paths; from existing ERMrest content; by
computing file checksums; or from file content. Handlers can also
compose or transform accumulated metadata, upload file data to Hatrac
or other HTTP-based object stores, or add metadata to ERMrest.

Another kind of IOBox agent can connect to an ODBC-compatible database
management server (e.g. Microsoft SQLServer, MySQL, etc.) and export a
set of user-specified query results in a portable serialization
format. A complementary agent consumes this serialized export and
loads content into an ERMrest catalog. These agents can be used for
periodic extraction and replication of content from local
lab-management systems to collaborative catalogs shared by multiple
labs.

\subsection{Condition-Action Processing}

Many user-agents can integrate with ERMrest. In our pilot projects, we
often see that there are natural parts of the ERM which already
reflect stateful conditions for condition-action process planning.
Idiomatic queries for expressing actionable conditions include:
entities with certain coded state or quality-assurance values;
entities with URL columns representing derived process results;
outer-joins showing entities which lack certain relationships to other
companion entities; or aggregated joins where certain heuristic
thresholds can be applied to related entity counts.  Arbitrary
mixtures of users and user-agents can participate in condition-action
processing. They can collectively monitor and advance the state of the
shared storage by pushing in new scientific context, new asset
references, and updates to asset metadata. This can include
integration of third-party processing pipelines.

A simple approach to detecting an actionable condition is through a
cron job that is scheduled to periodically query ERMrest and find data
to process. By executing a task for each discovered entity and
updating ERMrest to reflect new processing results, the cron job
advances the state of the environment in a way visible to external
observers and future iterations of itself. More sophisticated variants
can maintain error-tracking fields in the ERM to allow limited retry
or to flag failures for human intervention.

ERMrest can be configured to broadcast a {\em change notice} to an
AMQP message exchange after each catalog update, alerting interested
listeners that there may be new data. Rather than a stateless cron
job, a persistent agent can efficiently interleave queries to ERMrest
with blocking waits for these change notices. This allows low latency
response to new conditions, without wastefully polling the catalog
when content is quiescent.

\section{Applications}
\label{section:gpcr}

In Section~\ref{section:characteristics}, we characterized
the scientific data-management problem which we have generalized from a
number of data-oriented collaboration experiences. In this section, we
explore in more detail two of these applications and explain how
ERMrest has helped each. The GPCR and synaptomics problems represent
complex, distributed collaboration and early-phase exploratory
research, respectively.

\subsection{GPCR Consortium}

G-protein-coupled receptors (GPCRs) play a critical role in a wide
variety of human physiology and pathophysiological conditions. As a
drug target, GPCRs are highly valuable but mechanistically poorly
understood.  As computational methods are limited, the best method to
determine their structure is via X-Ray crystallography.  A challenge
is that the native form of the GPCR may not form a stable crystal, so
many slight mutations, called {\em constructs}, are designed and
evaluated.  For each construct, the protein is synthesized and
expressed in mass through bacteria. The protein is then purified out
of the bacteria, and evaluated for quality. If the protein is
crystallized, the crystal will then be evaluated against high energy
light sources, and finally have its structure determined by analysing
the diffraction patterns.  Various tests (assays) using techniques
such as flow cytometry, chromatography, and gel electrophoresis images
are used along the way to measure the quality, quantity and stability
of the resulting proteins.

\begin{figure}
  \includegraphics[width=\linewidth]{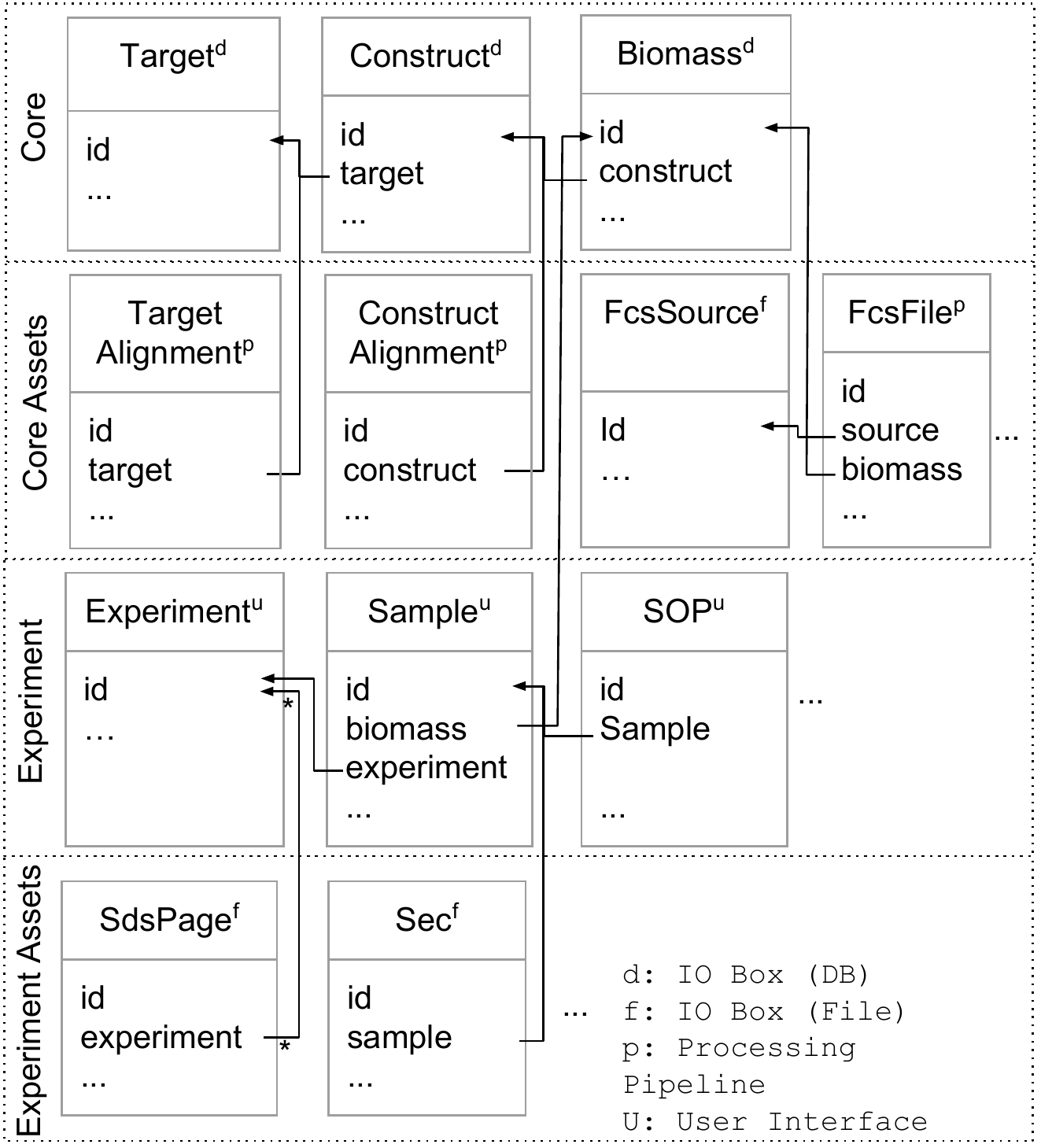}
  \caption{Select elements of GPCR catalog model. From top to bottom,
    four tiers of entities and relationships have been added in
    phases: {\em core} protein concepts; {\em core assets} including
    alignment data; {\em experiment} metadata; and most recently {\em
      experiment assets} are now being incorporated into the shared
    repository.}
  \label{fig:gpcr_erm}
\end{figure}

Partly due to the complexity of this process, the three dimensional
structures are known for only 154 out of the 826 different GPCRs.
Because of their importance, the GPCR Consortium, a distributed group
of academic and industrial partners, has been formed to systematically
evaluate a large number of GPCR structures. Typically, all of the
data associated with the process is managed in an ad hoc fashion,
often resulting in replicated experiments, lost data, and significant
investigator overheads.  Given the scale and distribution of the GPCR
Consortium effort, this conventional approach to data management would
pose a significant obstacle to the goals of the consortium.

Recognizing this challenge, the consortium has deployed a distributed
solution based on ERMrest with the objective of organizing all
consortium experiments and data.  The heart of our solution is an
evolving ERMrest catalog to track research products and pipeline
state. Figure~\ref{fig:gpcr_erm} illustrates essential parts of the
catalog ERM comprised of four tiers of entity types and relationships
developed in phases: A {\em core} ERM captures the domain of GPCR
targets, constructs, and biomasses; {\em core asset} metadata tracks
alignment and flow cytometry data files; {\em experiment} metadata
tracks additional processes; and most recently, {\em experiment asset}
metadata is beginning to track chromatography and electrophoresis data
files. Concurrent with these major phases of ERM expansion, we also
engaged the early users to review and refine the elements within each
tier.

All three academic sites have legacy databases with local construct
design and production information. We use IObox relational export and
import agents to maintain a harmonized, multi-site record of {\em
  core} entities in the shared catalog.  For file-like experimental
data such as flow cytometry ({\tt .FCS}), chromatography ({\tt .CDF}), or gel images
({\tt .JPG}), disk-monitoring IObox agents are deployed at each site. The
agents share the same general ingest pattern which is to check for
filename pattern, e.g. {\tt GPCRUSC20161013SDS1\_ABC123.jpg} for a
know experiment ID or {\tt UNKNOWN\_ABC123.jpg} for an unknown
experiment ID.  Files are automatically added to the shared object
store and cross-linked with entities in ERMrest based on detected
metadata.  The Chaise GUI is used for entering experiment
related metadata such as experiment design, associated samples,
purification protocols, or chemical composition. Assets which were
detected by IObox with unknown experiment context can be found by
users using Chaise and augmented with metadata after the fact.

Multiple processing and analysis pipelines are integrated as
condition-action sequences in cron jobs and persistent agents.  In
practice, a longer sequence is sometimes implemented as one script,
triggered asynchronously by the initial condition and performing
several idempotent steps.  This allows efficient recovery from partial
failures and performs complete tasks with fewer independent polling
agents, as intermediate failures are all recognized as ``incomplete''
states which continue to match the triggering condition.

\begin{figure}
  \includegraphics[width=\linewidth]{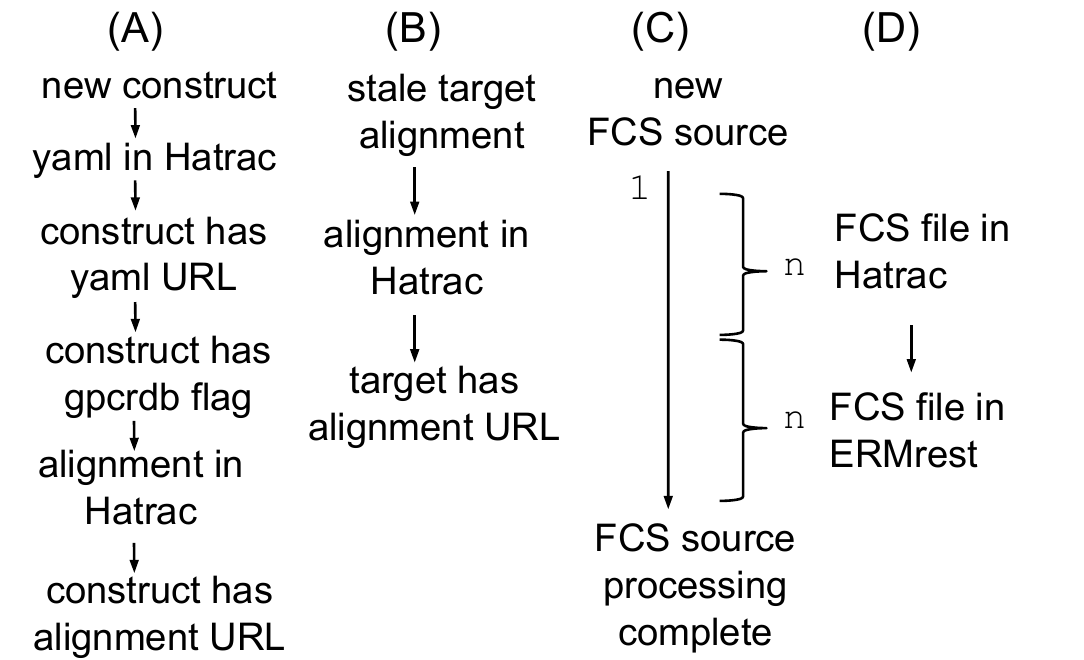}
  \caption{GPCR condition-action processing pipelines. Observable data
    states are depicted as labeled {\em conditions}, while processing
    {\em actions} are implied as arrows transitioning from one state
    to the next: A) a new construct is aligned using a third-party
    service, gpcrdb; B) an aggregate alignment is maintained for each
    target, tracking its most recent construct alignments; C) a
    multi-sample FCS source file is processed in bulk, generating
    idempotent checkpoints for D) single-sample FCS file.}
  \label{fig:gpcr_pipeline}
\end{figure}

In the core model, constructs arrive without alignment data. A chain
of condition-action steps, depicted in
Figure~\ref{fig:gpcr_pipeline}(a), show how the storage resources
transition from one state to another to fill in alignment data for a
construct.  Figure~\ref{fig:gpcr_pipeline}(b) depicts a coupled
condition-action sequence which prepares an aggregated alignment of
the target each time it receives a new construct
alignment.  We actually store hashes of alignments in the metadata
catalog to efficiently express express a ``stale target'' condition
as a polling metadata query.

The FCS processing pipeline nests a set of simple condition-action
sequences as idempotent sub-tasks in a larger bulk
action. Figure~\ref{fig:gpcr_pipeline}(c) depicts processing of a
multi-sample FCS file asset. The triggering condition is that the
corresponding FCS source is ``incomplete.'' The bulk action includes
expanding the file into constituent single-sample FCS files, each of
which is further processed and summarized. The bulk action can restart
multiple times, recognize already completed FCS file products, and
continue working until completion.

Chaise is the main user interface for consortium data.  We use a
number of Chaise-supported annotations on the catalog model to
customize data presentation, including adjustments to visible columns
for certain GUI contexts, rendering of URLs and attributes as links in
the web applications, and embedding content showing interactive
visualization elements or thumbnail images.  As described previously
in Section~\ref{section:characteristics}, GPCR data are subject to
differentiated access controls, enforced by Hatrac and ERMrest to
provide consistent policy enforcement for web browsers and any other
networked clients.

Prior to our involvement, all consortium data was managed in an ad hoc
fashion. Local site databases and networked storage appliances
provided essentially all-or-nothing access only to local content with
inconsistent file naming conventions, and most experiment metadata was
locked in scientists' personal notebooks.  Our solution has introduced
a consortium-wide catalog for capturing: core protein structure
results; experiment and assay status across the protein determination
workflow; and a viable system for sharing data according to the
consortium data-sharing policy. Integrated acquisition, processing,
and system state presentation GUIs have reduced the effort for users,
increased visibility into workflow status, helped increase data
quality, and freed up time for actual science.

The GPCR project has been in operation for 1.5 years. There are 928
targets including non-GPCRs, 23,240 constructs, 78,673 expressions,
and 51,752 FCS assets in the system. We continue to expand the
experiment data model tier to support protein purification and
crystallization process and asset tracking, including new asset types
such as chromatography data and gel electrophoresis images.

%

\subsection{Mapping the Dynamic Synaptome}

In 1894 Ramon Y Cajal first suggested that memories are formed by
changes in synaptic connections, a view that is widely held by
neuroscientists.  Over the last 50 years, studies examining the
behavior of individual synapses have not only demonstrated the
presence of synaptic plasticity, but also elucidated some of its major
properties and underlying mechanisms. However, it has not been
possible to address the question of how information is mapped onto
patterns of synapses across the brain, a prerequisite for
understanding the connection between brain structure and
behavior. What is needed are the means to measure the dynamic
synaptome: a map of the strength, location and polarity (excitatory or
inhibitory) of synapses in a living organism at different points in
time as they acquire new behaviors.  In this project, we seek to
address this fundamental question by creating a new high-throughput,
scalable method that enables direct observation and mapping of the
dynamic synaptome of the brain of a living organism.

Our paradigm closely couples three distinct technological advances: an
innovative method for creating recombinant probes to label
postsynaptic proteins in vivo, allowing synaptic strength to be
assessed in living animals without affecting neuronal function;
high-resolution, high-speed Selective Plane Illumination
Microscopy~(SPIM) to measure the 3D concentration of labeled
postsynaptic proteins across the brain; and a nimble and intuitive
platform and associated algorithms to drive high-throughput
acquisition of protein concentration maps from 3D images, to convert
these maps into a computable dynamic synaptome, and to follow the
characteristics of synapses over time.  We are developing new
techniques in all three areas, and have developed an initial
experiment protocol to explore elements of the pipeline, using 3D
brain images and behavioral videos.

\begin{figure}
  \includegraphics[width=\linewidth]{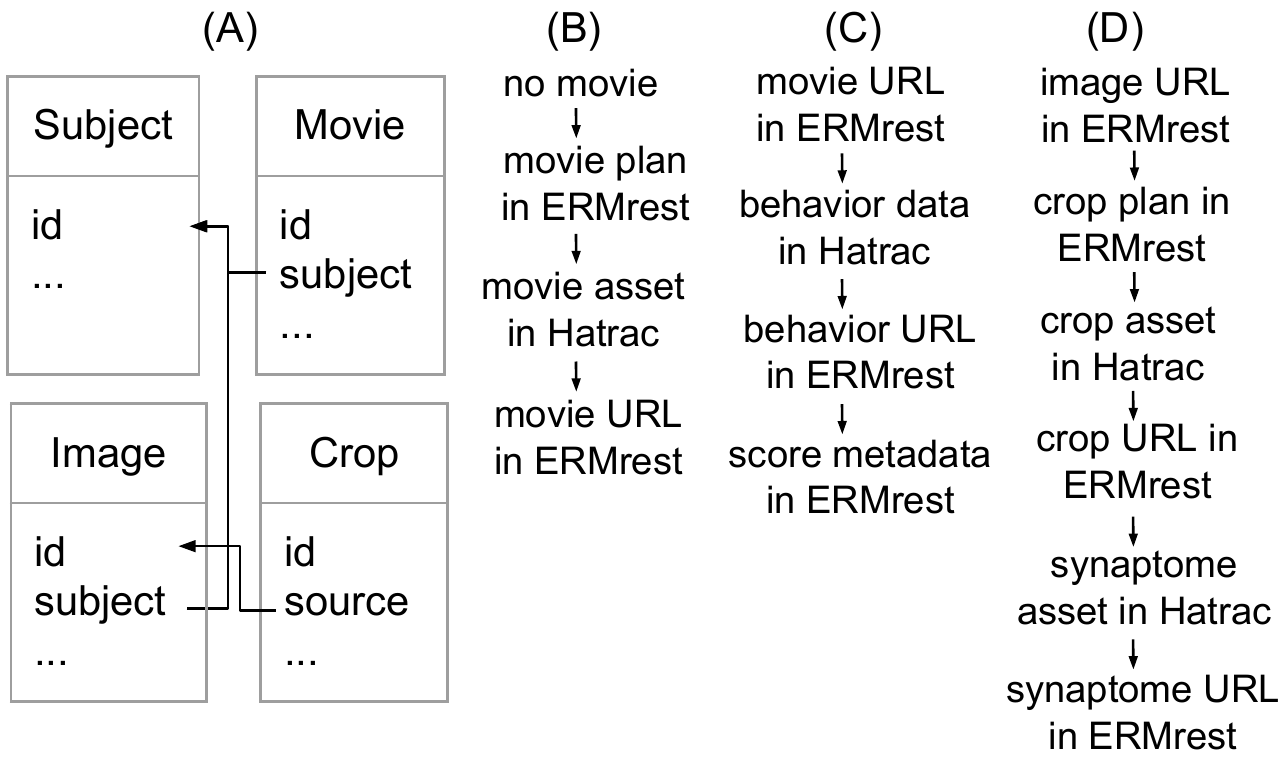}
  \caption{ Synaptome ERM (A) and condition-action sequence
    examples. The catalog represents pre-registered experiment plans
    entered by users prior to data-acquisition and processing: B)
    studies are entered into the catalog, acquired on an instrument,
    and handed to an IObox agent which stores the movie and adds the
    movie URL to the catalog; C) Movies are processed automatically
    and the resulting behavior data is likewise stored and registered,
    followed by a user reviewing the data and manually entering
    qualitative metadata; D) Eliding the image acquisition flow, a
    user operates a workstation to produce an image crop which is
    stored and registered via IObox, and then a user operates a
    workstation to detect and extract synaptome data from the image
    crop before it too is stored and registered.  }
  \label{fig:synapse}
\end{figure}

As for GPCR, we deployed an ERMrest catalog to track experiments and
data for this project. Figure~\ref{fig:synapse} illustrates essential
parts of the system: zebrafish larvae are tracked as {\em subjects};
SPIM {\em images} and behaviorial {\em movies} are collected from
instruments; SPIM image {\em crops} are extracted for targeted brain
regions; and analysis results are associated as asset URL and scoring
metadata on both cropped images and behavior movies. The behavior
movie analysis pipeline, depicted in Figure~\ref{fig:synapse}(C) is
handled with condition-action automation similar to those in GPCR.

In this early-stage, exploratory work, the image cropping and image
analyses Figure~\ref{fig:synapse}(D) are currently human-driven
processes. In practice an interactive user queries or browses the
catalog via the Chaise GUI to locate actionable data. They process
associated raw data assets using workstation-based, interactive tools,
and they submit cropped image and image analysis results as new files
via an IObox agent, similarly to raw instrument data acquisition.

We use the Chaise GUI to collect metadata in lieu of laboratory
notebooks or any other lab information management system. Modeled
entities correspond directly to planned lab activities, and SQL
triggers generate accession identifiers, which the user can easily
transcribe to physical materials in the lab and embed in the names of
files processed later by IObox agents. The preparation, imaging, and
behavioral study on each larva takes several hours, and the
experimentalist has plenty of time to record pertinent metadata.

Prior to this data-management system, our collaborators made use of
local lab filesystems in instruments and networked storage
appliances. We used a shared Dropbox folder to exchange early sample
data between the three teams. With a half-dozen active participants,
our project is a microcosm of the same data-management challenges seen
in our larger national and international projects. The file sizes,
file counts, and idiosyncratic naming conventions produced in
early-phase research threatened to overwhelm and interfere with
productivity. We find ourselves and our collaborators excited by the
impact our ERMrest-based tools are having.  By introducing these
data-sharing methods early, before we have even validated an
experiment pipeline or produced any science results, we are
accelerating the work; better organizing our data and activities to
understand the status of novel experimental techniques; and evolving a
data-management practice which, if we are successful, will also be at
the core of this new high-throughput laboratory method.

\section{Related Work}
\label{section:related}

ERMrest can be viewed as a metadata system to support publication.
Digital repository systems, such as DSpace~\cite{smith2003dspace} and
Globus Publish~\cite{chard2015globus}, provide object and data
collection level metadata, similar to the ERMrest-Hatrac
combination. Digital repositories are primarily concerned with
publication, as opposed to the discovery process itself where one's
understanding of the domain model may evolve considerably and hence
these systems have very limited metadata models (e.g. without
relationships) and don't support model evolution nor support easy
creation of multiple catalogs.

Other research has explored topics of integrated metadata catalogs
with key-value models~\cite{rajasekar2010irods}; distributed metadata
catalogs with key-value models~\cite{koblitz2008amga}; and distributed
relational database access underpinning metadata
catalogs~\cite{antonioletti2005design}.  However, these catalog
support a flat, per-asset description of data, and don't support the
structured models that ERMrest does, nor do they provide RESTful
interfaces.  Research on metadata catalogs has considered issues of
flexible modeling~\cite{deelman2004grid}, dynamic model generation and
integration~\cite{tuchinda2004artemis}, and incorporating semantic
representations~\cite{wang2009semantic} into metadata catalogs.
We differ from this work in focusing on ER modeling as being more
understandable by end users and integrating ERMrest into a RESTful web
services architecture.

SQLShare~\cite{howe2011database} is a system that has many elements in
common with ERMrest catalog including the concepts of schema evolution
and incremental refinement. However, SQLShare differs from our work in
several significant ways. It focuses on SQL as the primary interface
by which users interact and assumes that the data of interest is
primarily stored in the SQLShare database. As a
data-analysis platform, SQLShare treats data and derivations as a
directed graph of diverging data sets, while ERMrest focuses on a
convergent store for collaboratively-maintained, largely isochronous
metadata. In essence, SQLShare manages a set of individual tabular
data sets submitted either as data or as derivation queries, while
ERMrest provides RESTful interfaces to incrementally manage a mutable
store by modifying data content or adjusting its entity-relationship
model.

More closely related is HTSQL~\cite{HTSQL} which also maps a
relational query space to URLs, but focuses on query language rather
than data-management service interfaces. Both systems offer a form of
chained navigation with filters or other entity-selection notation to
retrieve relational data. HTSQL provides a query language meant for
humans, including conventional whitespace and punctuation-based
tokenization with rich projection and aggregation syntax, attempting
to compete with or replace SQL. By contrast, ERMrest offers queries
meant for web client machinery, with tokenization rules closely
aligned to URL encoding standards with navigation and filters as
elements in a hierarchical URL path notation. ERMrest focuses on
simpler classes of query relevant to implementing web-based data
browsers and condition-action agents. Unlike ERMrest, HTSQL does not
provide support for model introspection, model evolution, content
update, nor differentiated access control, where individual rows may
be visible to some clients but not others.

\section{Conclusion}
\label{section:conclusions}

We have summarized key design objectives and challenges faced in many
scientific data management and collaboration problems, and we have
introduced ERMrest, a web-based metadata catalog addressing these
problems. We have presented ERMrest goals, design, and
implementation, and described its ecosystem of client tools and
companion services.

We described two applications which have benefited from adoption of
ERMrest. The GPCR~Consortium is a complex, internationally
distributed collaboration with many active data-producers and
consumers, already having a mature set of experimental processes but a
mixture of different local laboratory data environments. The Mapping
the Dynamic Synaptome project is an early-phase, multidisciplinary
research project where the core science methodology is still being
invented, and no existing experiment management process was
known. Each project has successfully used ERMrest and its ecosystem of
tools to configure a project-specific data-sharing environment
which has accelerated their scientific work.

We have also described several possible areas for future work in
ERMrest, including: web-based management for fine-grained access
control policies within the ERM; web-based management of named views
and asynchronous query resources; and refinement of the data access
model to provide more compliant PATCH interfaces or other more RESTful
renderings of tabular data mutation. ERMrest and ERMrest-based tools
are open-source and are publically available on github
(github.com/informatics-isi-edu/ermrest).

\section*{Acknowledgment}

The authors would like to thank Serban Voinea for his contributions to ERMrest and IOBox development and Anoop Kumar and Alejendro Bugacov for their work on the GPCR project.  We would also like to acknowledge our GPCR collaborators Mike Hanson, Jeff Sui, and Raymond Stevens.
The work presented in this paper was funded by the National Institutes of Health  under awards 5U54EB020406,  1R01MH107238-01,  5U01DE024449 and 1U01DK107350 and by the GPCR Consortium.



\bibliographystyle{IEEEtran}
\bibliography{ermrest}

\begin{thebibliography}{10}
\providecommand{\url}[1]{#1}
\csname url@samestyle\endcsname
\providecommand{\newblock}{\relax}
\providecommand{\bibinfo}[2]{#2}
\providecommand{\BIBentrySTDinterwordspacing}{\spaceskip=0pt\relax}
\providecommand{\BIBentryALTinterwordstretchfactor}{4}
\providecommand{\BIBentryALTinterwordspacing}{\spaceskip=\fontdimen2\font plus
\BIBentryALTinterwordstretchfactor\fontdimen3\font minus
  \fontdimen4\font\relax}
\providecommand{\BIBforeignlanguage}[2]{{%
\expandafter\ifx\csname l@#1\endcsname\relax
\typeout{** WARNING: IEEEtran.bst: No hyphenation pattern has been}%
\typeout{** loaded for the language `#1'. Using the pattern for}%
\typeout{** the default language instead.}%
\else
\language=\csname l@#1\endcsname
\fi
#2}}
\providecommand{\BIBdecl}{\relax}
\BIBdecl
\renewcommand{\BIBentryALTinterwordstretchfactor}{4}

\bibitem{bell2009beyond}
G.~Bell, T.~Hey, and A.~Szalay, ``{Beyond the data deluge},'' \emph{Science},
  vol. 323, pp. 1297--1298, 2009.

\bibitem{kandel2012enterprise}
S.~Kandel \emph{et~al.}, ``Enterprise data analysis and visualization: An
  interview study,'' \emph{IEEE Transactions on Visualization and Computer
  Graphics}, vol.~18, no.~12, pp. 2917--2926, 2012.

\bibitem{Begley2013}
C.~G. Begley, ``{Six red flags for suspect work.}'' \emph{Nature}, vol. 497,
  no. 7450, pp. 433--4, may 2013.

\bibitem{borgman2012conundrum}
C.~L. Borgman, ``The conundrum of sharing research data,'' \emph{Journal of the
  American Society for Information Science and Technology}, vol.~63, no.~6, pp.
  1059--1078, 2012.

\bibitem{deriva-escience}
R.~Schuler, C.~Kesselman, and K.~Czjakowski, ``Accelerating data-driven
  discovery with scientific asset management,'' in \emph{IEEE 12th
  International Conference on eScience}.\hskip 1em plus 0.5em minus 0.4em\relax
  IEEE, 2016.

\bibitem{DOA}
------, ``Data centric discovery with a data-oriented architecture,'' in
  \emph{2015 Workshop on the Science of Cyberinfrastructure}, 2015.

\bibitem{fielding2002principled}
R.~T. Fielding and R.~N. Taylor, ``Principled design of the modern web
  architecture,'' \emph{ACM Transactions on Internet Technology (TOIT)},
  vol.~2, no.~2, pp. 115--150, 2002.

\bibitem{borgman2016}
C.~L. Borgman \emph{et~al.}, ``Data management in the long tail: Science,
  software, and service,'' \emph{International Journal of Digital Curation},
  vol.~11, no.~1, pp. 128--149, 2002.

\bibitem{RFC3986}
T.~Berners-Lee, R.~Fielding, and L.~Masinter, ``{RFC} 3986,'' \emph{Uniform
  Resource Identifier (URI): Generic Syntax}, 2005.

\bibitem{RFC4151}
\BIBentryALTinterwordspacing
T.~Kindberg and S.~Hawke, ``{The 'tag' URI Scheme},'' RFC 4151 (Informational),
  Internet Engineering Task Force, Oct. 2005. [Online]. Available:
  \url{http://www.ietf.org/rfc/rfc4151.txt}
\BIBentrySTDinterwordspacing

\bibitem{CommonMark}
S.~Leonard, ``Guidance on markdown: Design philosophies, stability strategies,
  and select registrations,'' Tech. Rep., 2016.

\bibitem{smith2003dspace}
M.~Smith \emph{et~al.}, ``Dspace: An open source dynamic digital repository,''
  Tech. Rep., 2003.

\bibitem{chard2015globus}
K.~Chard \emph{et~al.}, ``Globus data publication as a service: Lowering
  barriers to reproducible science,'' in \emph{e-Science (e-Science), 2015 IEEE
  11th International Conference on}.\hskip 1em plus 0.5em minus 0.4em\relax
  IEEE, 2015, pp. 401--410.

\bibitem{rajasekar2010irods}
A.~Rajasekar \emph{et~al.}, ``{iRODS} primer: integrated rule-oriented data
  system,'' \emph{Synthesis Lectures on Information Concepts, Retrieval, and
  Services}, vol.~2, no.~1, pp. 1--143, 2010.

\bibitem{koblitz2008amga}
B.~Koblitz, N.~Santos, and V.~Pose, ``The {AMGA} metadata service,''
  \emph{Journal of Grid Computing}, vol.~6, no.~1, pp. 61--76, 2008.

\bibitem{antonioletti2005design}
M.~Antonioletti \emph{et~al.}, ``The design and implementation of grid database
  services in ogsa-dai,'' \emph{Concurrency and Computation: Practice and
  Experience}, vol.~17, no. 2-4, pp. 357--376, 2005.

\bibitem{deelman2004grid}
E.~Deelman \emph{et~al.}, ``Grid-based metadata services,'' in \emph{Scientific
  and Statistical Database Management, 2004. Proceedings. 16th International
  Conference on}.\hskip 1em plus 0.5em minus 0.4em\relax IEEE, 2004, pp.
  393--402.

\bibitem{tuchinda2004artemis}
R.~Tuchinda \emph{et~al.}, ``Artemis: Integrating scientific data on the
  grid,'' in \emph{AAAI}, 2004, pp. 892--899.

\bibitem{wang2009semantic}
X.~Wang \emph{et~al.}, ``Semantic enabled metadata management in petashare,''
  \emph{International Journal of Grid and Utility Computing}, vol.~1, no.~4,
  pp. 275--286, 2009.

\bibitem{howe2011database}
B.~Howe \emph{et~al.}, ``Database-as-a-service for long-tail science,'' in
  \emph{International Conference on Scientific and Statistical Database
  Management}.\hskip 1em plus 0.5em minus 0.4em\relax Springer, 2011, pp.
  480--489.

\bibitem{HTSQL}
\BIBentryALTinterwordspacing
(2012) {HTSQL} web site. [Online]. Available: \url{http://www.htsql.org}
\BIBentrySTDinterwordspacing

\end{thebibliography}

\end{document}